\documentclass[12pt]{iopart}
\usepackage{iopams}
\usepackage{setstack}

\usepackage{graphicx}
\usepackage{dcolumn}
\usepackage{bm}
\usepackage{float}
\usepackage[english]{babel} 

\usepackage[numbers,sort&compress]{natbib}

\usepackage[usenames,dvipsnames]{xcolor}
\usepackage[colorlinks=true,citecolor=Cerulean,linkcolor=RubineRed,urlcolor=Cerulean]{hyperref}

\begin{document}

\title{Barriers to Macroscopic Superfluidity and Insulation in a 2D Aubry-Andr\'e Model}
\author{Dean Johnstone$^1$, Patrik \"{O}hberg$^1$ and Callum W. Duncan$^{2}$}
\address{$^1$ SUPA, Institute of Photonics and Quantum Sciences,
	Heriot-Watt University, Edinburgh, EH14 4AS, UK}
\address{$^2$ Department of Physics, SUPA and University of Strathclyde, Glasgow G4 0NG, United Kingdom}
\ead{dj79@hw.ac.uk, p.ohberg@hw.ac.uk, callum.duncan@strath.ac.uk}
\date{\today}

\begin{abstract}
We study the ground state phases of interacting bosons in the presence of a 2D Aubry-Andr\'e potential. By using a a mean-field percolation analysis, we focus on several superlattice and quasicrystalline regimes of the 2D Aubry-Andr\'e model, including generalisations that account for a tilting or skewing of the potential. We show that barriers to the onset of macroscopic phases naturally arise from weakly modulated domains in the 2D Aubry-Andr\'e model. This leads to the formation of mixed phases, in which the macroscopic properties are dominated by a minority of the system. The phase diagrams then exhibit substantially different features when compared against crystalline systems, including a lobe-like or wave-like appearance of the Bose glass, sharp extrusions and extended domains with weak percolation. By studying the 2D Aubry-Andr\'e model across multiple regimes, we have shown that the unique properties of mixed phases are not distinct to a small set of parameters.
\end{abstract}

\section{Introduction}
The Bose Glass (BG) is a famous example of a quantum phase that stabilises a coexistence of insulating and SuperFluid (SF) ground states on local scales \cite{Fisher1989}. While the BG is viewed as macroscopically insulating, local SF domains will mean that the phase is compressible, unlike a Mott-Insulator (MI). Furthermore, these SF domains do not percolate, leading to the absence of macroscopic phase coherence \cite{greiner2002,PhysRevB.80.214519,barman2013understanding,Niederle_2013}. The BG is perhaps most commonly associated to disordered, crystalline systems. In this case, short-range random disorder is known to introduce atomic localisation for both interacting \cite{PhysRev.115.2,PhysRevLett.10.159,PhysRevB.21.2366,PhysRevB.37.325} and non-interacting systems \cite{PhysRev.109.1492,RevModPhys.50.191}. The disorder averaged phase is of particular interest in these models, as no single disorder realisation should dominate the overall physics. In other words, the local fluctuations do not play an important role in the underlying phase transitions. For this study, however, we wish to study the properties of models that are not random. In this scenario, local variations of any order parameters are far more important than they would be for disordered systems. This can allow for the formation of mixed phases, where the macroscopic properties of a phase are dominated by a minority of the system.

The system we study is based on a 2D extension of the quasiperiodic Aubry-Andr\'e (AA) model with on-site interactions. The AA model and its generalisations have been widely studied in the single-particle picture \cite{aubry1980analyticity}, and are known to host self-dualities \cite{mott1987mobility,PhysRevB.83.075105,PhysRevB.102.024205}, novel dynamical properties \cite{PhysRevB.99.094203,PhysRevB.97.174206,PhysRevB.99.224204} and topological phases \cite{PhysRevLett.110.180403,PhysRevA.93.062101,PhysRevB.101.020201}. While many studies have been conducted in 1D, various 2D extensions of the AA model have also found similar properties \cite{PhysRevB.99.054211,PhysRevB.101.014205,PhysRevLett.116.140401}. For quasiperiodic systems, we have a distinct scenario in which we have both short-range disorder and long-range order present in the system. Importantly, however, this short range disorder is not random, and the systems structure is normally characterised by long-range self-similarity. Furthermore, this order can naturally lead to the formation of barriers to macroscopic phases.
	
Barriers to macroscopic phases manifest as weakly modulated domains in the 2D AA potential, which can form in both quasiperiodic and periodic limits. These domains can then stabilise the mixed phases on the lattice. This includes regimes where a small SF domain of a few sites can percolate through the system and support macroscopic superfluidity, despite a majority of the phase possessing insulating characteristics. We can also observe the opposite scenario, where small, percolating domains with insulating behaviour will block the onset of macroscopic superfluidity.

Recently, the BG has been theoretically observed for some 2D quasicrystalline systems through a mean-field analysis of tight-binding models \cite{johnstone2021meanfield} and quantum Monte Carlo studies of continuous systems \cite{PhysRevLett.126.110401}. To-date, the mixed phases have not been studied in detail, which is the central aim of this work. We will explore the local nature of ground state phases for the many-body 2D AA model.  We achieve this through a Gutzwiller mean-field analysis, based on percolation methods for inhomogeneous systems \cite{PhysRevLett.75.4075,PhysRevB.80.214519,PhysRevB.85.020501,barman2013understanding,Niederle_2013,Nabi_2016,PhysRevA.98.023628,johnstone2021meanfield}.

Here, we will present our results for the 2D AA model as follows. First, in Sec.~\ref{sc_2}, we will define the 2D AA model that we work with in this study, including a discussion on the mean-field percolation methods and order parameters used to determine different macroscopic and mixed phases. Using these definitions, we then discuss the structure of mixed phases in Sec.~\ref{sc_str}, and show several examples of the mixed phases on the lattice for different forms of the underlying potential. Finally, in Sec.~\ref{sc_5}, we present full ground state phase diagrams for the system in different parameter regimes, before ending with our conclusions in Sec.~\ref{sc_6}.

\begin{figure}[t]
	\centering
	\includegraphics[width=0.75\linewidth]{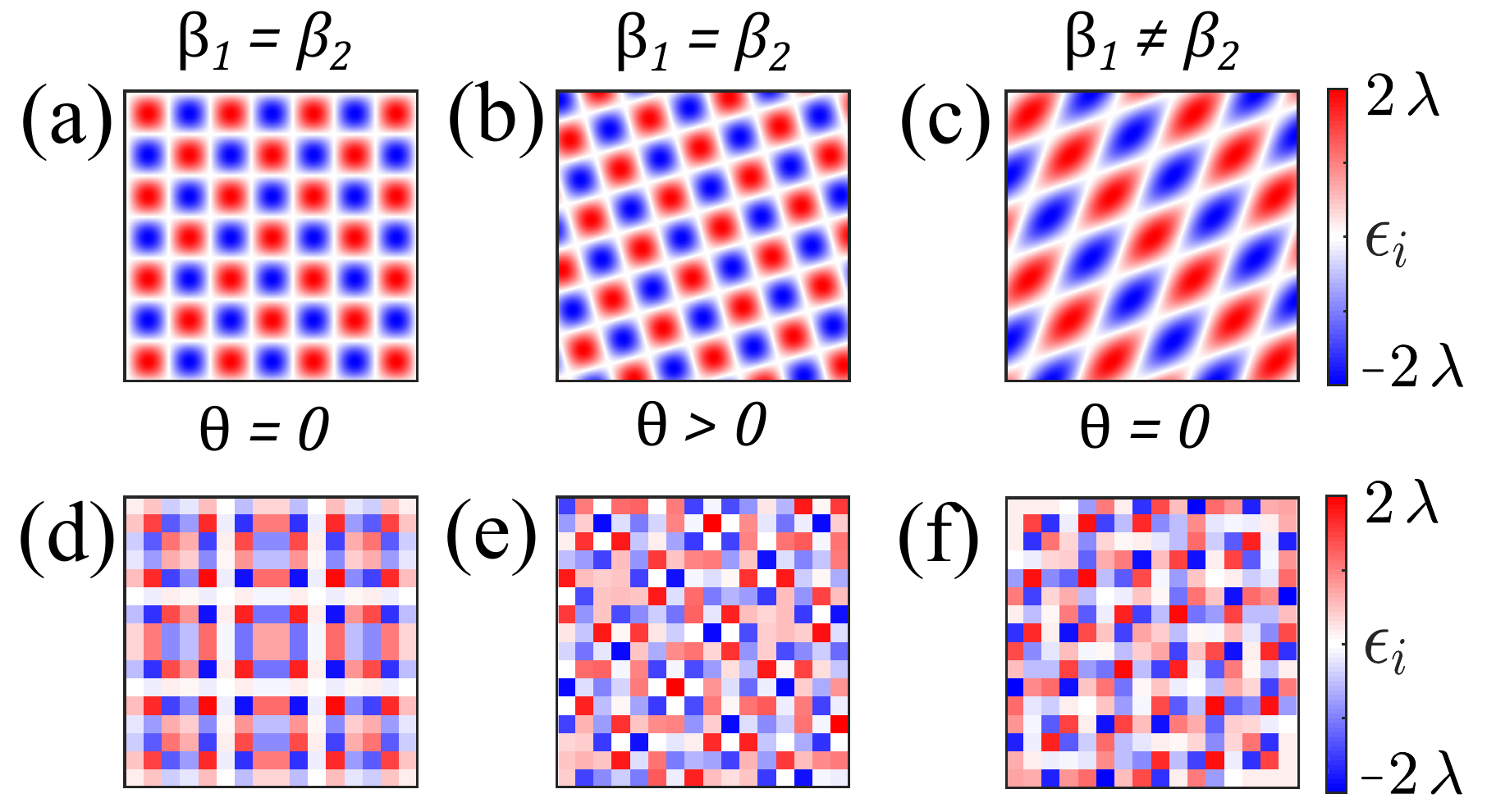}
	\caption{Plots of the 2D AA potential for the three different regimes used in this study, including (a,d) equal wavenumbers without a rotation $\theta$ of the potential, (b,e) equal wavenumbers with rotation and (c,f) unequal wavenumbers without rotation. The first row of figures (a-c) denote the continuum limits of Eq.~\eref{eq_aaEps}, whereas the bottom row (d-f) shows the 2D AA potential on a discrete lattice.}
	\label{figure_eps}
\end{figure}

\section{Mean-field 2D Aubry-Andr\'e model} \label{sc_2}

\subsection{2D Aubry-Andr\'e Bose-Hubbard Model}

In the following, we will consider an inhomogeneous Bose-Hubbard model on a 2D square lattice with unit spacing and Hamiltonian
\begin{equation}	\label{eq_bhm}
\hat{H} = \frac{U}{2} \sum_i^N \hat{n}_i (\hat{n}_i -1) + \sum_i^N (\epsilon_i - \mu) \hat{n}_i - J\sum_{\langle i,j\rangle} \hat{b}^\dagger_i \hat{b}_j ,
\end{equation}
where $N$ is the total number of lattice sites, $U$ is the on-site interaction strength, $\epsilon_i$ is an on-site energy, $J$ is the tunnelling coefficient, $\langle i,j\rangle$ denotes the sum over nearest-neighbours, $\mu$ is the chemical potential, $ \hat{b}_i(\hat{b}^\dagger_i) $ are the bosonic annihilation(creation) operators and $\hat{n}_i$ is the number operator. We will consider the $\epsilon_i$ to be distributed according to a 2D AA quasiperiodic potential of
\begin{equation} 	\label{eq_aaEps}
\epsilon_i = -\lambda \Big[ \cos\big(2 \pi \beta_1 (x + y)\big) + \cos\big(2 \pi \beta_2 (x - y)\big) \Big],
\end{equation}
with $\lambda$ denoting the modulation strength, $\beta_{1,2}$ are the wavenumbers and $x, \, y$ are the 2D spatial coordinates of the lattice. Throughout this work, we will consider a $N = 99 \times 99$ lattice.

The Bose-Hubbard model can be efficiently realised with cold gases trapped in optical lattices \cite{Jaksch1998,greiner2002}. The presence of quasiperiodic order, such as the 2D AA potentials considered here, can then be introduced with separate optical potentials \cite{PhysRevLett.91.080403,roati2008anderson,PhysRevLett.98.130404,Deng2009,PhysRevA.91.043632}. In 2D, it has been demonstrated that this will reproduce known results with speckle potentials when random phase fluctuations are introduced \cite{PhysRevLett.95.070401,pasienski2010disordered,PhysRevLett.126.110401}, including the formation of glass states \cite{PhysRevLett.98.130404,PhysRevA.77.063605} and many-body localisation \cite{Schreiber842,PhysRevB.87.134202}. It has also been shown that 1D quasiperiodic models can support many-body localisation \cite{PhysRevB.87.134202,PhysRevLett.122.170403} and a variety of intermediate phases before thermalisation \cite{PhysRevB.102.195142,PhysRevResearch.1.032039,PhysRevLett.122.170403,PhysRevLett.126.080602}. Furthermore, other studies have found exotic ground states \cite{PhysRevA.81.023626,PhysRevLett.125.060401} and localisation transitions \cite{PhysRevB.101.174203,PhysRevA.78.023628,Deng2009,PhysRevLett.115.180401,mastropietro2017localization}.

If the wavenumbers in Eq.~\eref{eq_aaEps} are irrational, they will be incommensurate with the lattice spacing. The distribution of $\epsilon_i$ will then be quasicrystalline, and contain long-range order. On the other hand, if the wavenumbers are rational, we will instead have a commensurate, crystalline distribution of $\epsilon_i$ on the lattice. Throughout this study, we will refer to the potentials with irrational wavenumbers as quasicrystalline distributions and rational wavenumbers as superlattice distributions. In Fig.~\ref{figure_eps}, we plot visualisations of the 2D AA potential for the three different regimes used in this study. Starting with equal wavenumbers in Figs.~\ref{figure_eps}(a,d), we can see the appearance of weakly modulated lines with $\epsilon_i=0$, as observed in previous studies \cite{PhysRevB.101.014205,johnstone2021meanfield}. These lines are predictable from the form of Eq.~\eref{eq_aaEps} as
\begin{equation} 	\label{eq_modLn}
d(Y) = Y - \frac{k}{4\beta} \approx 0,
\end{equation}
where k is an odd integer, $\beta$ is a fixed wavenumber when $\beta_1=\beta_2$ and $Y$ can be either the $x$ or $y$ coordinate. If $d(Y)$ is sufficiently close to zero, then the corresponding row/column of sites will be weakly modulated. In the single particle picture, it is known that these weakly modulated lines can destabilise the mobility edge that is typically seen in AA models, and can even support ballistic transport \cite{PhysRevB.101.014205}.

In this work, we will also study other quasicrystalline distributions of the potential that, in general, no longer stabilise precise lines of weak modulation that match the geometry of the underlying lattice. Of particular relevance to experimental protocols is the consideration of tilted quasicrystalline potentials. For example, this would amount to the rotation of some bichromatic quasiperiodic potential on top of an optical lattice, which could describe alignment errors. We will consider a rotation of the 2D AA potential, which transforms the spatial coordinates of Eq.~\eref{eq_aaEps} according to
\begin{equation}
\pmatrix{x \cr y} \rightarrow \pmatrix{x\cos{\theta} - y\sin{\theta} \cr x\sin{\theta} + y\cos{\theta}},
\end{equation}
where $\theta$ denotes an anti-clockwise rotation angle from the $x$-axis.  We show an example of a tilted 2D AA potential in Figs.~\ref{figure_eps}(b,e). 

Finally, we will also consider the case of unequal wavenumbers. There is of course many different choices one can make for the wavenumbers, and we shall consider a particular set in order to demonstrate the rich physics that is present in the many-body 2D AA model. We will therefore consider the following parametrisation for the wavenumbers
\begin{equation}
\beta_1 = \sin{\phi}, \, \beta_2 = \cos{\phi},
\end{equation}
where $\phi$ is an effective skew ``angle" between $\beta_1$ and $\beta_2$. This ensures at least one irrational wavenumber for $\phi>0^{\circ}$. Given the reflectional symmetry about $\phi=45^{\circ}$, we therefore consider a range of $\phi$ between $45^{\circ}$ and $0^{\circ}$. A skewed 2D AA potential is depicted in Figs.~\ref{figure_eps}(c,f), which shows a similar structure to that of the tilted potentials.

\subsection{Gutzwiller Mean-Field}

In order to study the many-body properties of the 2D AA model, we consider a mean-field percolation analysis. First, we decouple correlators according to a Gutzwiller ansatz \cite{Rokhsar1991,Krauth1992,Gutzwiller1965}
\begin{equation}
\hat{b}^\dagger_i \hat{b}_j = \hat{b}^\dagger_i \varphi_j + \hat{b}_j \varphi_i - \varphi_i \varphi_j ,
\end{equation}
where $\varphi_i = \langle \hat{b}_i \rangle$ is the mean-field order parameter at site $i$, which is taken to be real without loss of generality for the considered model. By substituting the above relation into Eq.~\eref{eq_bhm}, the Hamiltonian for the inhomogeneous Bose-Hubbard model can be written in the local number basis for each site as
\begin{equation} \label{eq_locH}
\hat{H}_i = \frac{U}{2} \hat{n}_{i} (\hat{n}_{i} - 1) + (\epsilon_i - \mu) \hat{n}_{i} - \, J(\hat{b}_{i} + \hat{b}_{i}^\dagger)\sum\limits_{\langle i,j \rangle} \varphi_j .
\end{equation}
The ground state for the entire system can then be found by diagonalising Eq.~\eref{eq_locH} for each site and converging. During this process, the order parameters are updated according to the local number basis
\begin{equation} \label{eq_ord_p1}
\varphi_i = \langle \hat{b}_{i} \rangle = \sum\limits_{n=0}^z \sqrt{n}f_{n}^{(i)} f_{n-1}^{*(i)} ,
\end{equation}
where $z$ is the maximum number of particles per site and $f^{(i)}_n$ are elements of the lowest energy eigenvector for site $i$. Given an initial set of order parameters $\varphi_i$, this process can be repeated in a self-consistent manner, until convergence to the true ground state. We can also calculate the local density for each site
\begin{equation} \label{eq_ord_p2}
\rho_i = \langle \hat{n}_{i} \rangle = \sum\limits_{n=0}^z n | f_{n}^{(i)} |^2 .
\end{equation}
As a practical note, $z$ should in principle be infinite for bosonic systems, but it is set to a finite value for numerical purposes. The ground state will converge with increasing $z$. We take $z=10$ for the results presented here, which is well into the region of convergence.

\subsection{Percolation and Mixed Phases}

The simplest case of the homogeneous Bose-Hubbard model with on-site interactions normally permits the existence of two unique ground state phases; the MI and SF. In the context of mean-field theory, the presence of this macroscopic order can be based on whether the average order parameter $\bar{\varphi}$ is finite. This can be captured with the correlation fraction $\mathcal{F}$ of a state, which we define as
\begin{equation}
\mathcal{F} = \frac{N_{\varphi}}{N},
\end{equation}
where $N_{\varphi}$ is the total number of sites with finite $\varphi_i$ and $N$ is the total number of sites. In other words, $\mathcal{F}=0$ for the MI and $\mathcal{F}\sim1$ for the SF.

If the model has disorder, then a BG is expected to prevent a direct MI to SF transition \cite{PhysRevB.40.546}. Unlike the SF, a BG no longer supports macroscopic phase coherence, and is hence also an insulating phase. However, the BG will have small, isolated SF domains, which means that the phase is compressible, unlike the MI \cite{PhysRevB.40.546}. We can identify a BG-SF transition based on the percolation of sites with a local SF character, as has been done in prior works on percolation based methods \cite{PhysRevLett.75.4075,PhysRevB.85.020501,Niederle_2013,Nabi_2016}. To do this, we define a percolation probability $\mathcal{P}$ as
\begin{equation} \label{eq_PPerc}
\mathcal{P} = \frac{N_{span}}{N_{\varphi}},
\end{equation}
where $N_{span}$ is the number of sites in a percolating cluster. For the MI and BG, we now have $\mathcal{P}=0$, whereas for the SF $\mathcal{P}>0$. The application of a percolation analysis with conventional mean-field approaches has been shown to produce comparative results to those obtained by quantum Monte-Carlo for disordered systems on a square lattice \cite{PhysRevB.80.214519,PhysRevLett.99.050403,PhysRevLett.107.185301,Niederle_2013}.

For the 2D AA model, there is the possibility that short-range, non-random disorder can result in phases that are dominated by local properties of the system. An example of this could be a thin line of SF sites percolating across an otherwise insulating system. This state, while being for the most part insulating, could be considered to host a macroscopic SF. The macroscopic properties of the ground state are then dominated by the local properties of comparatively few sites, e.g. for the $N = 99 \times 99$ system we consider, a single percolating line of SF sites could be supported by as little as $1\%$ of the sites. We will refer to ground states dominated by local properties as \textit{mixed phases}.

We will characterise mixed phases by considering two local discrete functions that measure the phase at each site and its surroundings. These will be referred to as locality functions. First, we define the insulating locality function $S_{i}^{\mathrm{MI}}$, which is $1$ if a MI site (integer density and zero $\varphi_i$) is surrounded by MI sites across each of its bonds, and $0$ in all other scenarios. The complimentary function $1 - S_{i}^{\mathrm{MI}}$ is therefore $1$ if a MI site has at least one neighbouring SF site. Similarly, we can also define the SF locality function $S_{i}^{\mathrm{SF}}$, which is $1$ if a SF site (non-zero $\varphi_i$) is surrounded by SF sites across each of its bonds, and $0$ otherwise. The complimentary function $1 - S_{i}^{\mathrm{SF}}$ is then $1$ if a site has at least one neighbouring MI site. By looking at the average values of these distributions, $S^{\mathrm{MI}}$ and $S^{\mathrm{SF}}$, we end up with a measure between $0$ and $1$ which is more sensitive to the local structure and fluctuations of clusters.

\begin{figure}[t]
	\centering
	\includegraphics[width=0.98\linewidth]{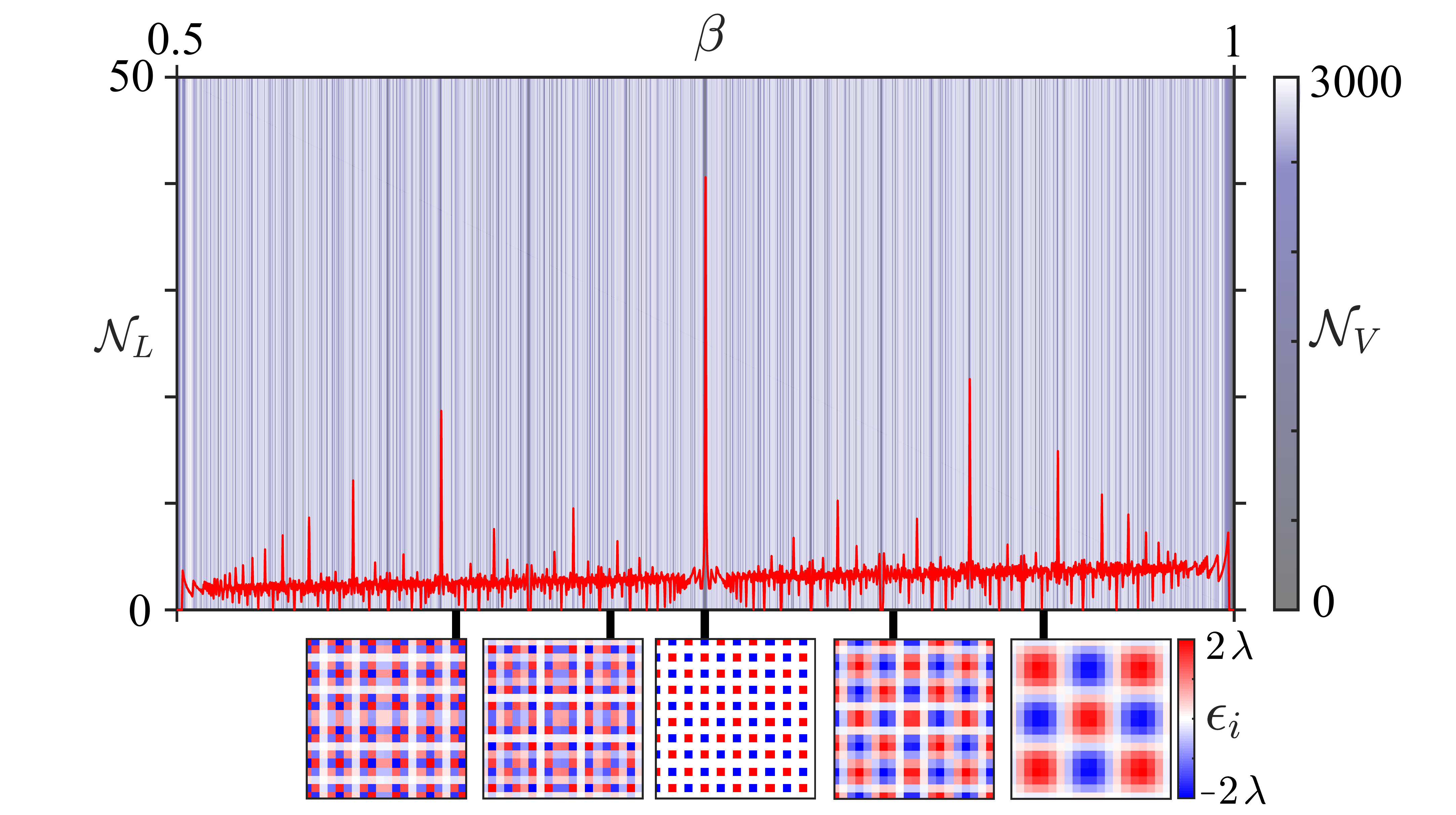}
	\caption{The red line shows the number of weakly modulated lines $\mathcal{N}_L$ as a function of $\beta$. The curve has been smoothed to better illustrate the most prominent features. We also include a background plot which shows the number of unique values (up to $3$ significant figures) $\mathcal{N}_V$ in the energy potential as a function of $\beta$. Several examples of the AA potential are plotted at the bottom of the figure for visualisation purposes. As expected, $\mathcal{N}_V$ is minimised around rational $\beta$, but quickly inflates as we approach quasicrystalline distributions. $\mathcal{N}_L$, on the other hand, can either be peaked or minimised to zero around rational $\beta$. However, $\mathcal{N}_L$ will remain largely finite across a large range of irrational $\beta$.}
	\label{figure_uniqueV}
\end{figure}

From this, we can first define a weak SuperFluid (wSF) phase, which has a majority of sites being insulating, but has a small percolating SF domain. This thin SF cluster acts as a barrier to macroscopic insulation within the system. In other words, more than $50 \%$ of sites will still have a MI character. Note, at the $50 \%$ threshold of MI to SF sites, $S^{\mathrm{MI}}$ and $S^{\mathrm{SF}}$ can change significantly, depending on the local distribution of clusters. For example, if the system is composed of two large clusters of MI and SF character on the square lattice, the maximum values of the locality functions will be 
\begin{equation}
S^{\mathrm{MI}} = S^{\mathrm{SF}} = \frac{N-2\sqrt{N}}{2N},
\end{equation}
which is $0.4899$ for a $N=99 \times 99$ lattice. On the other hand, if there is checkerboard pattern of MI and SF sites, each site will be surrounded by neighbours with different phases, resulting in $S^{\mathrm{MI}} = S^{\mathrm{SF}} = 0$. A similar result also follows if the rows/columns on the lattice possess an oscillating MI/SF character. The wSF ground state phase supports macroscopic superfluidity, and will therefore be characterised by $S^{\mathrm{SF}} \approx 0$, $S^{\mathrm{MI}} \ge 0$, and $\mathcal{P}>0$.
 
The opposite scenario is that of a weak Bose Glass (wBG), which has a majority of sites being in the SF phase, but has small MI domains which prevents a SF percolating through the full system, meaning the state lacks macroscopic phase coherence. This ground state will then be characterised by $S^{\mathrm{SF}} \ge 0$,  $S^{\mathrm{MI}} \approx 0$ and $\mathcal{P}=0$. 

Finally, we note that for the macroscopic SF and MI phases, the corresponding locality functions will converge towards $S^{\mathrm{SF}} = 1$ and $S^{\mathrm{MI}} = 1$ respectively.

		
		
%


\section{Weak Modulation Lines}

From Eq.~\eref{eq_modLn}, we can expect weakly modulated domains to frequently appear in the 2D AA model for a large range of wavenumbers $\beta_{1,2}$. These domains can act as barriers to macroscopic percolation, and will therefore influence the formation of mixed phases. In order to show that this is indeed the case, we will consider the relation between the wavenumbers and the total number of weakly modulated lines. First, we take a weakly modulated line to be defined when the difference function $d(Y) < 10^{-2}$ for a given row/column, and will consider the limit of $\beta_1 = \beta_2 \equiv \beta$. For each row/column of the potential, we then count the total number of weakly modulated lines $\mathcal{N}_L$ that fall below this threshold as a function of $\beta$ for a $99 \times 99$ lattice, shown by the line plot in Fig.~\ref{figure_uniqueV}. Due to the quasiperiodic nature of the potential, $\mathcal{N}_L$ will rapidly oscillate with small changes in $\beta$. Importantly, however, there are many regions at irrational $\beta$ in which $\mathcal{N}_L$ is finite. We also colour regions of the plot according to the number of unique values $\mathcal{N}_V$ (to $3$ significant figures) in the 2D AA potential. As expected, the number of unique values will be minimised around rational wavenumbers, such as $\beta=[2/3,\, 3/4,\, 4/5 \, ...]$, and the resulting potential will take a superlattice form.

\section{Example Mixed Phases} \label{sc_str}

\begin{figure}[t]
	\centering
	\includegraphics[width=1.0\linewidth]{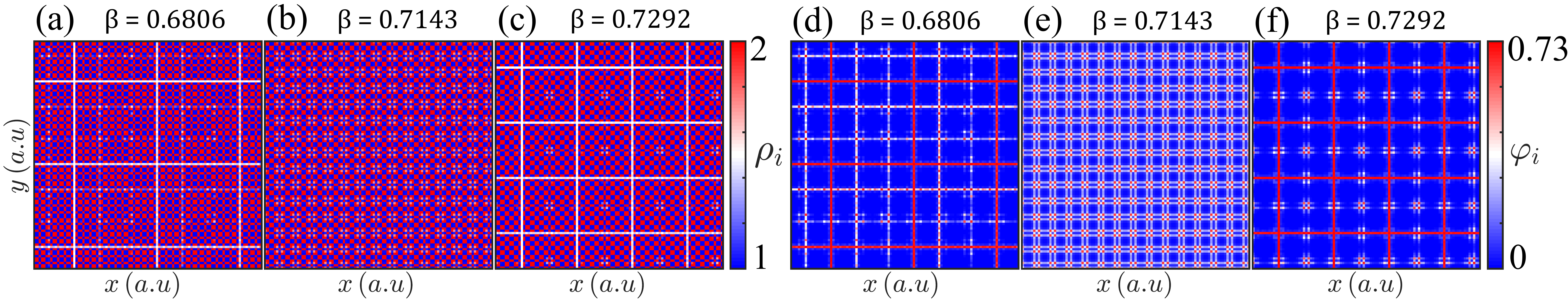}
	\caption{Plots of the order parameters at $\mu/U=1.0$ and $\lambda/U=0.15$, for different wavenumbers $\beta$ and tunnelling $J/U$. Each column corresponds to a fixed wavenumber and tunnelling $J/U$ of (a,d) $J/U=0.004$, (b,e) $J/U=0.008$ and (c,f) $J/U=0.004$, where we plot the local (a-c) density $\rho_i$ and (d-f) correlation $\varphi_i$ order parameters. The wavenumber in (b,e) corresponds to a rational wavenumber of $\beta=5/7$, whereas the other wavenumbers are irrational. In the quasicrystalline limits, we see percolating lines of densities around $\rho_i=3/2$, which indicate an enhanced transition of the local SF clusters.}
	\label{figure_ord_map_B1}
\end{figure}

\begin{figure}[t]
	\centering
	\includegraphics[width=1.0\linewidth]{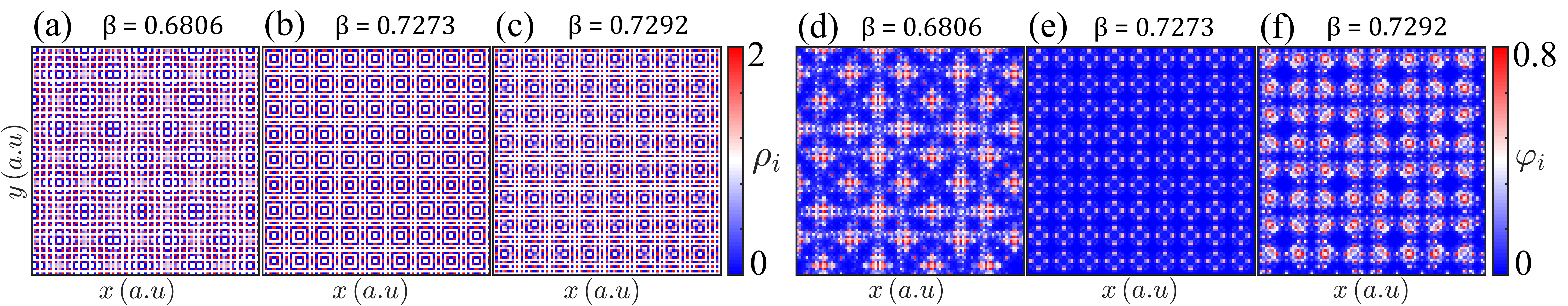}
	\caption{Plots of the order parameters at $\mu/U=0.39$ and $\lambda/U=0.525$, for different wavenumbers $\beta$ and tunnelling $J/U$. Each column corresponds to a fixed wavenumber and tunnelling $J/U$ of (a,d) $J/U=0.027$, (b,e) $J/U=0.02$ and (c,f) $J/U=0.025$, where we plot the local (a-c) density $\rho_i$ and (d-f) correlation $\varphi_i$ order parameters. The wavenumber in (b,e) corresponds to a rational wavenumber of $\beta=8/11$, whereas the other wavenumbers are irrational. In each case, we have BG/wBG phases near the onset of percolation.}
	\label{figure_ord_map_B2}
\end{figure}

In order to better understand the influence of mixed phases, we will now consider the structure of wSF and wBG phases on the lattice. First, we will discuss the case of equal wavenumbers $\beta_1 = \beta_2 \equiv \beta$, which results in the stabilisation of weak modulation lines. For this case, we can observe both the wSF and wBG phase in Figs.~\ref{figure_ord_map_B1} and~\ref{figure_ord_map_B2}, where we have plotted the local density $\rho_i$ and order parameter $\varphi_i$. In Fig.~\ref{figure_ord_map_B1}, we can clearly see the presence of a wSF phase, in which we have narrow percolating lines of SF states with non-zero $\varphi$. The wBG phase is a little more subtle, but it can be seen in Fig.~\ref{figure_ord_map_B2} that it is possible to have large numbers of sites with non-zero $\varphi$ that do not percolate through the full system. Fig.~\ref{figure_ord_map_B2}(d) is a particularly clear example of this.

\begin{figure}[t]
	\centering
	\includegraphics[width=1.0\linewidth]{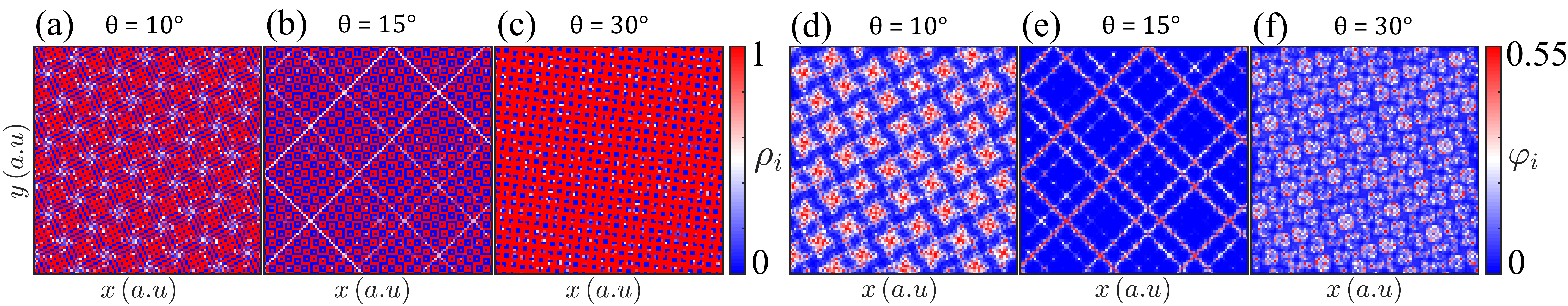}
	\caption{Plots of the order parameters for $\lambda/U=0.35$, with different tilt angles $\theta$. Each column corresponds to a fixed tilt angle, with the tunnelling and chemical potential given as (a,d) $J/U=0.032$; $\mu/U=0.1$, (b,e) $J/U=0.022$; $\mu/U=0.0$ and (c,f) $J/U=0.026$; $\mu/U=0.29$. We also denote the local (a-c) density $\rho_i$ and (d-f) correlation $\varphi_i$ order parameters.}
	\label{figure_ord_map_R}
\end{figure}

Next, we now turn to the more general cases of the tilted and skewed potentials. As we have shown in Fig.~\ref{figure_eps}, it is also possible for both the tilted and skewed potentials to exhibit similar lines of weak modulation. These domains will of course no longer form as precise vertical or horizontal lines on the square lattice, but rather as correlated patches of weak modulation. The strength of modulations across these patches will be more significant than what is typically observed on the weak modulation lines, which will therefore affect the stability and structure of mixed phases. In Fig.~\ref{figure_ord_map_R}, we show examples of the mixed phases for the tilted potentials. Here, we can clearly observe the formation of a wSF phase across diagonal lines of weak modulation in Fig.~\ref{figure_ord_map_R}(e). Furthermore, the wBG can also form for certain regimes in Figs.~\ref{figure_ord_map_R}(d) and (f). These mixed phases possess smaller clusters of MI sites across specific diagonal/zig-zag patterns on the lattice, which will again prevent the onset of SF percolation.

\begin{figure}[t]
	\centering
	\includegraphics[width=1.0\linewidth]{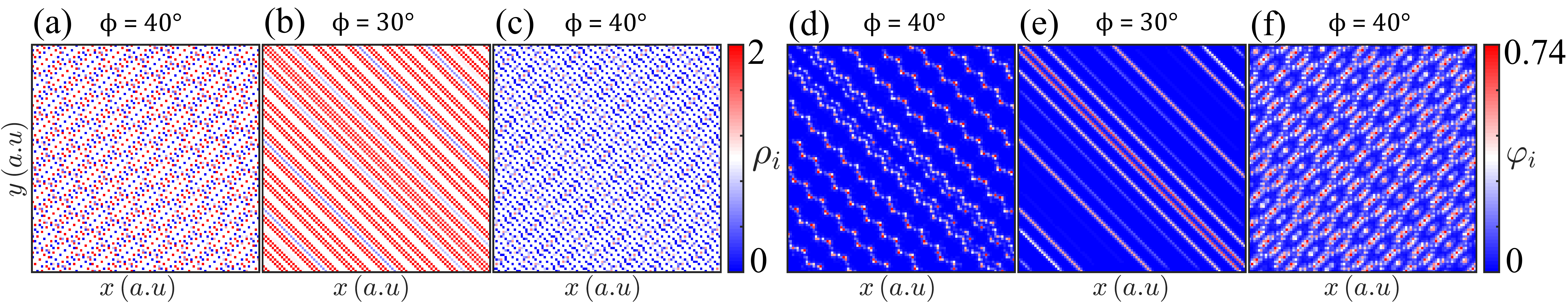}
	\caption{Plots of the order parameters for $\lambda/U=0.35$, with different skew angles $\phi$. Each column corresponds to a fixed skew angle, with the tunnelling and chemical potential given as (a,d) $J/U=0.015$; $\mu/U=0.5$, (b,e) $J/U=0.01$; $\mu/U=0.7$ and (c,f) $J/U=0.027$; $\mu/U=0.3$. We also denote the local (a-c) density $\rho_i$ and (d-f) correlation $\varphi_i$ order parameters.}
	\label{figure_ord_map_L}
\end{figure}

We also consider several example phases for the skewed potentials in Fig.~\ref{figure_ord_map_L}. For these potentials, the wSF can sometimes appear in a more trivial manner across diagonal regions, with some examples in Figs.~\ref{figure_ord_map_L}(d) and (e). The wSF phase in Fig.~\ref{figure_ord_map_L}(e) is a special case in which SF clusters only form across diagonal lines, giving rise to a 1D structure of the order parameters. Finally, the wBG can also form with interesting local properties and patterns of MI clusters, with an example in Fig.~\ref{figure_ord_map_L}(f).

\section{Many-Body Phase Diagrams} \label{sc_5}

\subsection{$\beta_1 = \beta_2$}	\label{sc_5_1}

\subsubsection{Mixed Phases}

\begin{figure*}[t]
	\centering
	\includegraphics[width=0.99\linewidth]{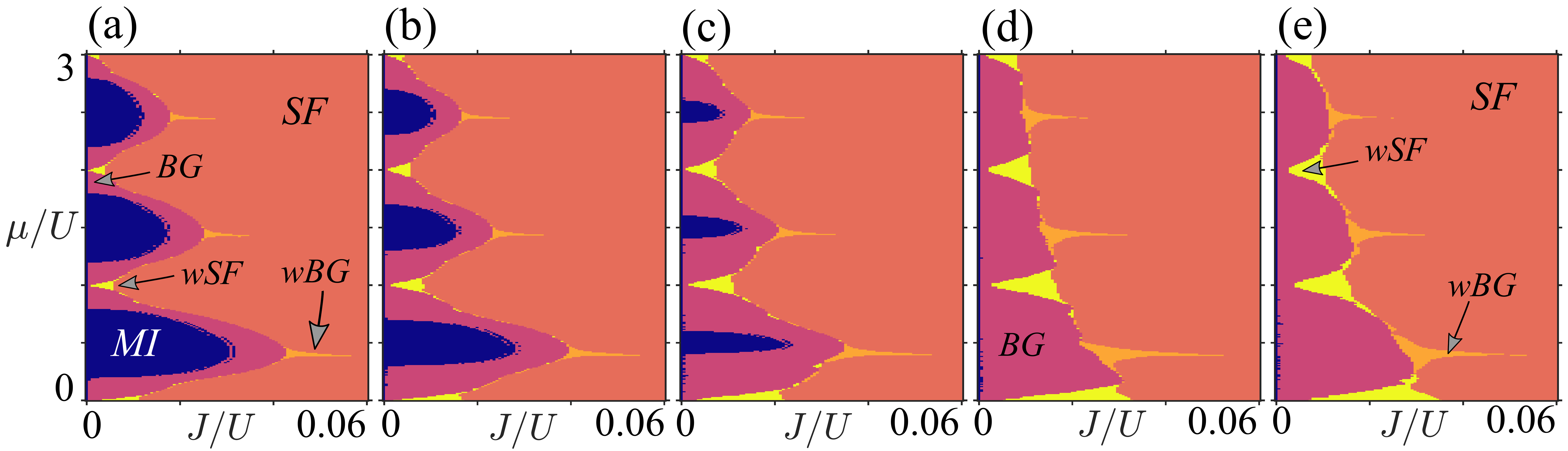}
	\caption{Phase diagrams of the 2D AA model for fixed modulation strengths $\lambda/U$. Here, we consider (a) $\lambda/U=0.1$, (b) $\lambda/U=0.15$, (c) $\lambda/U=0.2$, (d) $\lambda/U=0.35$ and (e) $\lambda/U=0.525$. As we increase $\lambda/U$, the MI lobes are slowly reduced in extent, leaving behind larger regions of the BG phase. At strong $\lambda/U$, the extruding features appearing from the BG lobes are still persistent. Furthermore, at integer $\mu/U$, the onset of the SF phase occurs at a small $J/U < 0.01$, with the wSF denoting regions of weak percolation.}
	\label{figure_phMap_md}
\end{figure*}

\begin{figure}[t]
	\centering
	\includegraphics[width=0.9\linewidth]{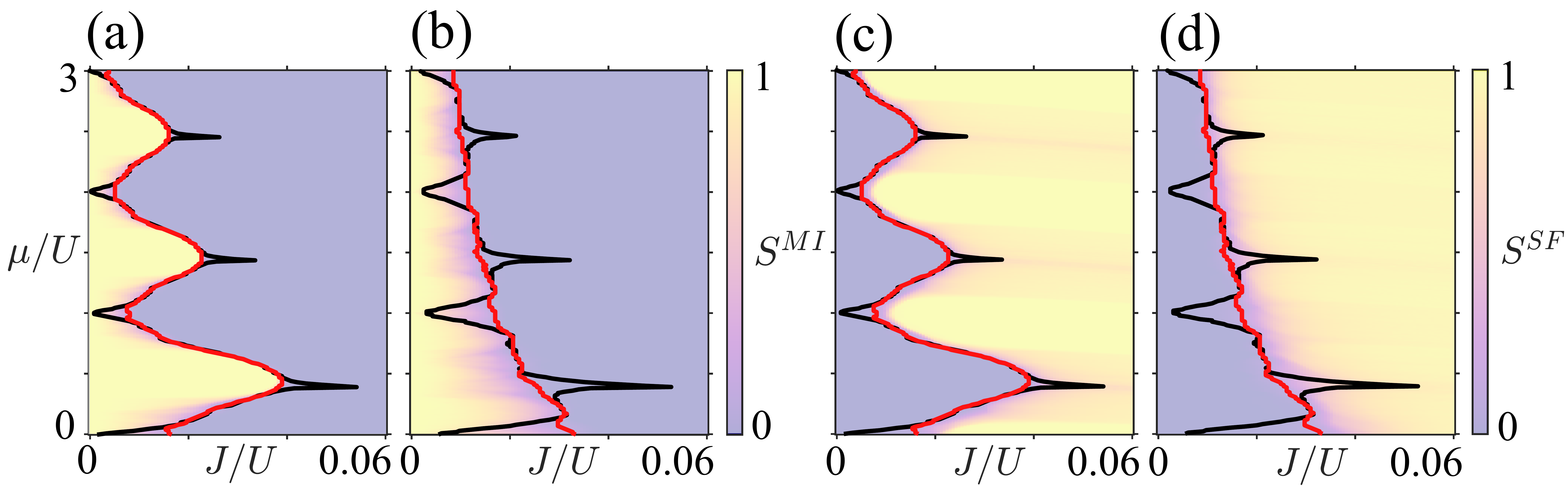}
	\caption{Plots of the (a,b) $S^{MI}$ and (c,d) $S^{SF}$ local measures for the characterisation of mixed phases. Here, we consider the two phase diagrams in Figs.~\ref{figure_phMap_md}(b,d), which correspond to modulation strengths of (a,c) $\lambda/U=0.15$ and (b,d) $\lambda/U=0.35$. The black line in each plot is the percolation transition, while the red line indicates when SF clusters account for more than $50 \%$ of the overall state.}
	\label{figure_phMap_MI_SF_nn}
\end{figure}

In Fig.~\ref{figure_phMap_md}, we plot phase diagrams for the system as function of $J/U$ and $\mu/U$ for $\beta=1/\sqrt{2}$, with different $\lambda/U$. We segment the ground state phase into SF, MI, and BG domains, as has been previously considered. However, we can now confirm the presence of wSF and wBG domains for the 2D AA model. The average of the discrete locality functions, $S^{\mathrm{MI}}$ and $S^{\mathrm{SF}}$, are shown in Fig.~\ref{figure_phMap_MI_SF_nn} for the phase diagrams of Fig.~\ref{figure_phMap_md}. We also plot additional boundaries on these phase diagrams, with the red line indicating when SF sites account for more than $50 \%$ of the overall phase, and the black line for the onset of SF percolation. The average locality functions show that there are domains where $S^{\mathrm{MI}}$ and $S^{\mathrm{SF}}$ are non-zero but not unity, meaning that all sites are not yet surrounded by the same local phase. From these average locality functions, we can define the wSF and wBG phases as previously discussed. The presence of the wSF and wBG phases is far clearer in the case of larger $\lambda/U$, shown by the extended transition in the locality functions in Fig.~\ref{figure_phMap_MI_SF_nn}(b,d). This shows a direct relation between the strength of the 2D AA potential and the presence of the mixed phases, which is highlighted further by the growth of the wSF and wBG domains in Fig.~\ref{figure_phMap_md} for increasing $\lambda/U$. 

A particularly noteworthy feature in Fig.~\ref{figure_phMap_md} is the presence of lobe-like BG domains, with peculiar extruding features of a wBG character. In Ref.~\cite{johnstone2021meanfield}, these features were speculated to form due to the presence of weakly modulated lines. We can now confirm -- through the measurement of the wBG -- that this is indeed the case. The sharp protruding features of the wBG are due to the support of an insulating phase from the weakly modulated lines, and such a structure would most likely be destroyed in a fully disordered system. We also find that the wSF is stabilised by the weak modulation lines near integer $\mu/U$. This is due to small SF clusters forming and percolating throughout the system, but with support from relatively few lattice sites.

\subsubsection{Mixed Phases and Critical Points}

Due to the inhomogeneous nature of the 2D AA model, calculating the full ground state phase diagrams over large ranges of $\beta$ would be inefficient. For the sections that follow, we will instead focus on the properties which will be most $\beta$ dependent, i.e. the mixed phases. Varying $\beta$ in general can only impact two of the properties of the system: (i) $\beta$ can be tuned to a quantity commensurate with the underlying lattice, therefore, away from being a quasicrystal, or (ii) $\beta$ can alter the appearance/location of the weak modulation lines that support the mixed phases. In order to study the effects of both scenarios, we will consider the behaviour of the BG-SF critical points at different $\mu/U$ for varying $\beta$. We choose two critical points, which correspond to the lobe-like behaviour of the BG and stabilisation of a wSF at an integer $\mu/U=1$, and the extrusion of the wBG at $\mu/U=0.39$. It is at these points of the ground state phase diagram that the presence of mixed phases is particularly pronounced. In \ref{sc_ap1}, we also consider full phase diagrams towards a flat, commensurate limit of $\beta=1$.

\begin{figure}[t]
	\centering
	\includegraphics[width=0.6\linewidth]{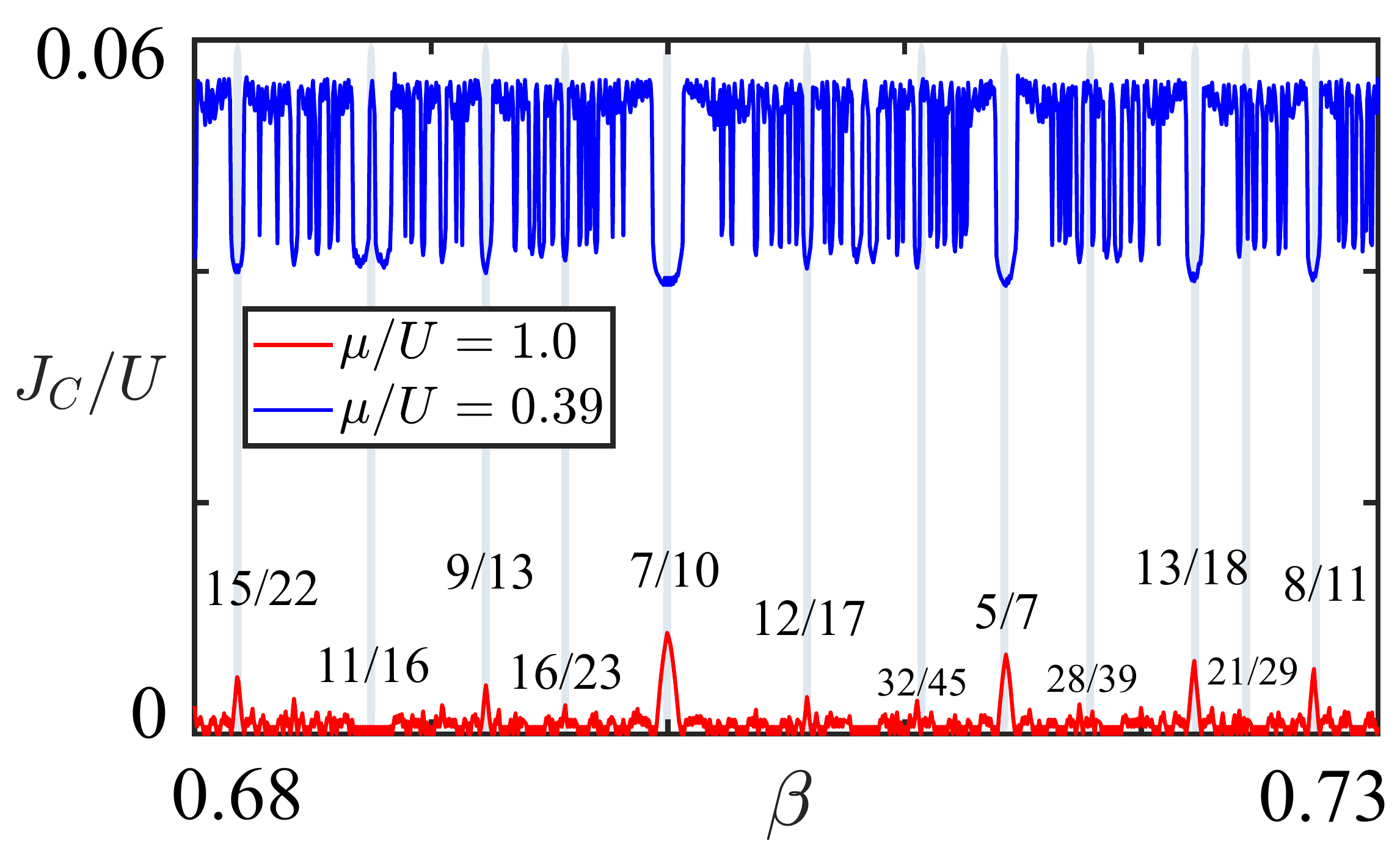}
	\caption{Plots of the critical behaviour at fixed chemical potentials for $\lambda/U=0.15$, showing the $J_C/U$ BG-SF critical point. For certain $\beta$, we also show the rational fraction above $J_C/U$ peaks and on the $\beta$ grids as vertical lines. The structure of local SF clusters can be divided into two distinct regimes based on whether or not $J_C/U$ is maximised or minimised for each $\mu/U$. These then directly correspond to superlattice and quasicrystalline realisations of the AA potential.}
	\label{figure_critJ_B}
\end{figure}

In Fig.~\ref{figure_critJ_B}, we plot the BG-SF critical points as a function of $\beta$. Starting with the red line, where $\mu/U=1$, we can immediately see that the critical points show little variation in this interval, with most $J_C/U \sim 10^{-3}$. The largest fluctuations occur for the superlattice limits of the 2D AA potential, which destabilise the lobe-like structure of the BG. This behaviour can be linked to the underlying structure of mixed phases, as observed in Fig.~\ref{figure_ord_map_B1}. For quasicrystalline distributions with irrational $\beta$, the local SF clusters account for a very small fraction of the overall phase. Despite this small fraction, the local SF clusters still percolate due to their formation across weakly modulated lines. This is in contrast to what is seen for superlattice potentials with rational $\beta$, where the SF clusters generally account for a much larger fraction. The structure of local SF clusters can take a more crystalline form in these limits, and may even introduce larger fluctuations to the weakly modulated lines. This short-range order will influence the percolation of local SF clusters, and therefore destabilise the lobe-like structure observed on the phase diagrams.

We also plot the BG-SF critical points for $\mu/U=0.39$, which is the blue line in Fig.~\ref{figure_critJ_B}, corresponding to the extruding feature for the wBG. Here, we see that this particular feature is quite sensitive to smaller changes in $\beta$. The positions of the local minima and maxima of $J_C/U$ shows several analogous properties to what was observed previously. For quasicrystalline $\beta$, we know that the phase will possess many SF sites, as per Fig.~\ref{figure_ord_map_B2}. Small MI domains across weakly modulated regions will, however, prevent percolation until a large $J_C/U$. When considering superlattice potentials, the onset of short-range order can remove such regions in the system, leading to a reduction in the $J_C/U$ critical point.

Depending on the rational fraction of $\beta$ in Fig.~\ref{figure_critJ_B}, we can also link the structure of mixed phases to the range of short-range order. Generally speaking, for the smaller numerators/denominators, the critical point will be significantly shifted from what is seen at irrational $\beta$. The reason for this is due to the absence of long-range variations in the on-site potential. As the numerators/denominators are enlarged towards irrational limits, we introduce long-range variations to the on-site energies $\epsilon_i$, and hence we observe less dramatic shifts in critical behaviour. These properties effectively mimic what is seen in Fig.~\ref{figure_uniqueV} for the number of weakly modulated lines and unique values. When $\beta$ is irrational, long-range variations in energy can allow for the appearance of very weakly modulated lines in Eq.~\eref{eq_modLn} at specific rows and columns. If $\beta$ is rational, then it is possible that the condition in Eq.~\eref{eq_modLn} may not be satisfied for any row or column, leading to the absence of barriers to macroscopic percolation.

\subsubsection{Density Waves from Superlattice Potentials}

\begin{figure}[t]
	\centering
	\includegraphics[width=0.85\linewidth]{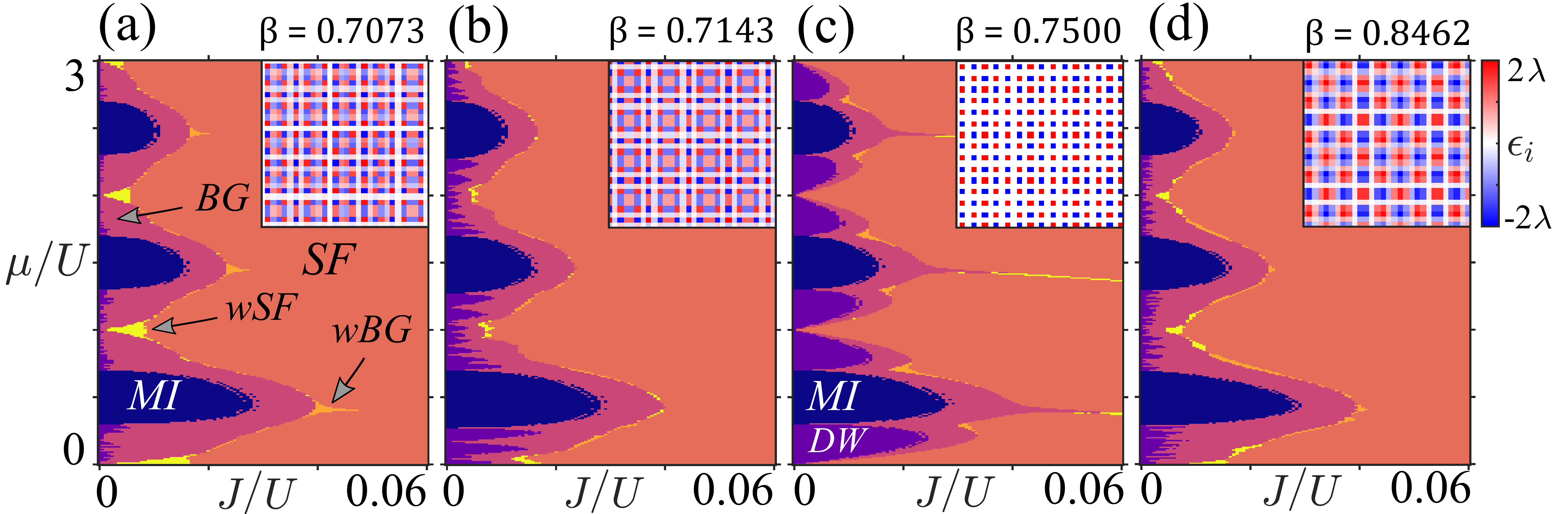}
	\caption{Phase diagrams of the 2D AA model for rational wavenumbers $\beta$ and a fixed modulation strength of $\lambda/U=0.15$. The inset figures represent small portions of the 2D AA potential. We also express the wavenumber to 4 decimal places, for comparison with Fig.~\ref{figure_uniqueV}. The exact wavenumbers are given as (a) $\beta=\frac{29}{41}$, (b) $\beta=\frac{5}{7}$, (c) $\beta=\frac{3}{4}$ and (d) $\beta=\frac{11}{13}$. For superlattice realisations of the potential with smaller fluctuations in energy, wSF and wBG phases will appear less frequently on the phase diagrams. Furthermore, it is also possible to stabilise macroscopic DW order at finite $J/U$.}
	\label{figure_phMap_beta}
\end{figure}

We now focus on the case of superlattice potentials when $\beta$ is rational, which is an interesting limit of the 2D AA model. We plot the ground state phase diagrams for a number of different rational wavenumbers in Fig.~\ref{figure_phMap_beta}. With the superlattice potentials, it is still possible to observe wSF and wBG phases in certain regimes, but their domain is severely reduced compared to the case of quasicrystalline potentials.

Interestingly, for some of the superlattice potentials, we can also observe the formation of macroscopic DW phases at finite tunnelling strengths, as shown in Fig.~\ref{figure_phMap_beta}. Similar to the MI, the DWs are macroscopically insulating and possess no transport. They will, however, contain non-uniform densities. Each distinct DW lobe therefore corresponds to a different non-integer filling of the lattice. The size and structure of DW lobes and intermediate wBG/wSF phases depends on the rational wavenumber and extent of long-range fluctuations. If the numerator/denominator of $\beta$ is small, there will only be a few distinct on-site energies in the 2D AA potential. The number of distinct DW lobes that appears on the phase diagram will then also be small, as seen in Figs.~\ref{figure_phMap_beta}(b,c). Furthermore, the width of these DW states is comparable to the size of the MI lobes in $J/U$. For the larger numerators/denominators of $\beta$ in Figs.~\ref{figure_phMap_beta}(a,d), fluctuations in on-site energies will become more pronounced, and the number of unique DW states will increase. The width of each DW lobe in $J/U$ will decrease, however, indicating a stronger sensitivity of these states to particle number fluctuations.

\begin{figure}[t]
	\centering
	\includegraphics[width=0.5\linewidth]{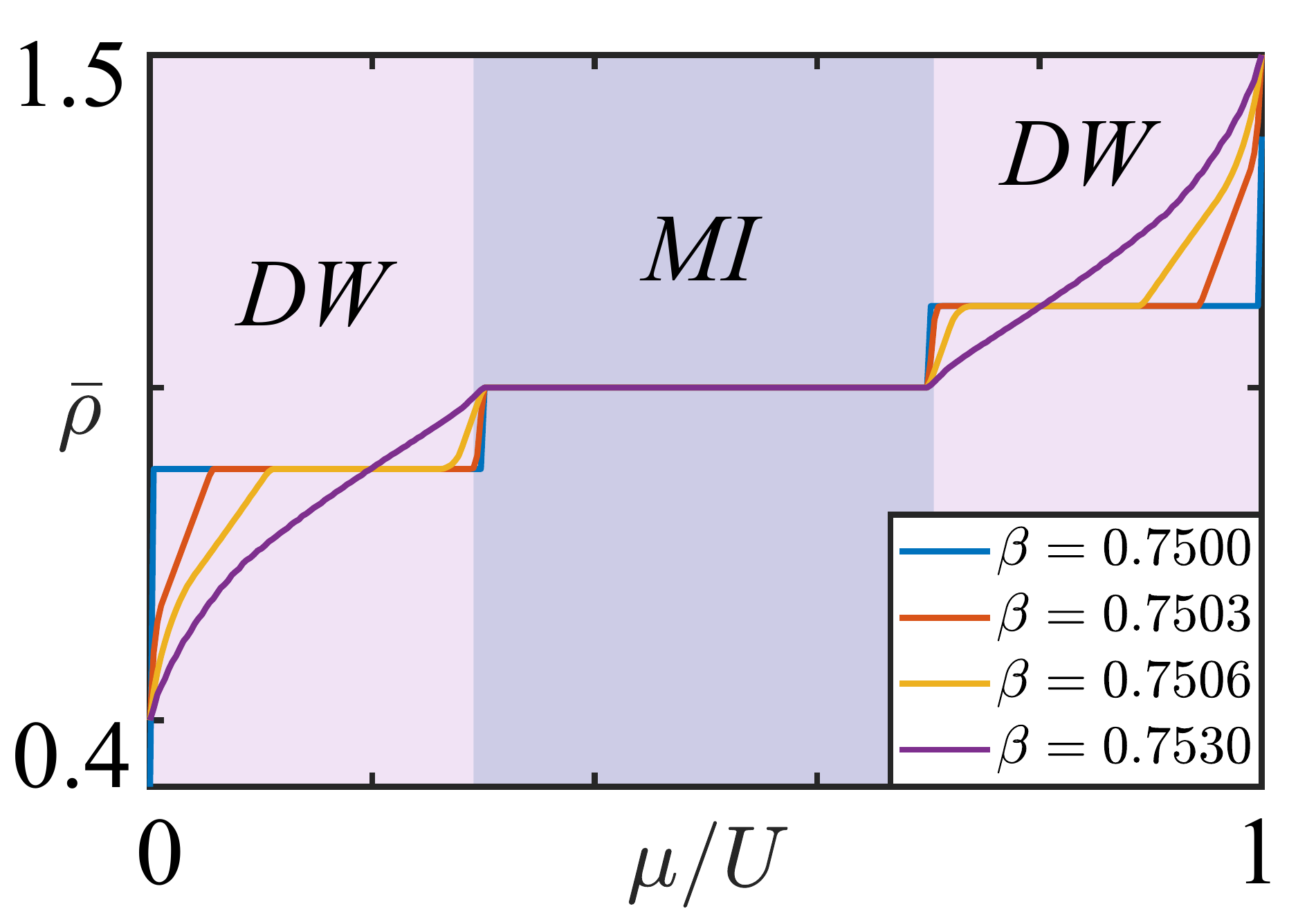}
	\caption{Average density as a function of $\mu/U$ around $\beta=3/4$, when $J/U=0$ and $\lambda/U=0.15$. All states are insulating for these parameters. We observe that the average density plateaus in the DW domains become unstable as we move further away from the rational limit in $\beta$. This is a consequence of the number of unique values in the energy potential increasing.}
	\label{figure_dw_vary}
\end{figure}

To better illustrate the stability of macroscopic DW order in the 2D AA model, we plot the average density in Fig.~\ref{figure_dw_vary}, as a function of $\mu/U$ around a rational wavenumber of $\beta=3/4$, when $J/U=0$ and $\lambda/U=0.15$. Here, we see that when $\beta$ is varied between rational and irrational limits, the properties of the DW states begin to drastically change. Notably, we observe that long plateaus of average density no longer form in the DW regions when $\beta$ starts to become incommensurate with the lattice. In other words, the total number of distinct DW states has significantly increased, but their width in $\mu/U$ is vanishingly small, giving rise to an almost continuous behaviour of $\bar{\rho}$ at $J/U=0$. It is important to note that away from the rational limit of $\beta=3/4$, the number of unique values in the potential will quickly inflate, as seen in Fig.~\ref{figure_uniqueV}. The small width of each DW in $\mu/U$ implies that the system will be very sensitive to particle number fluctuations, and hence why even small tunnelling strengths can immediately destroy DW order in these regimes. Similar properties are also observed around other rational $\beta$.

\subsection{Tilted Potential with $\beta_1 = \beta_2$} \label{sc_5_2}

\subsubsection{Ground State Phase Diagrams}

\begin{figure*}[t]
	\centering
	\includegraphics[width=0.99\linewidth]{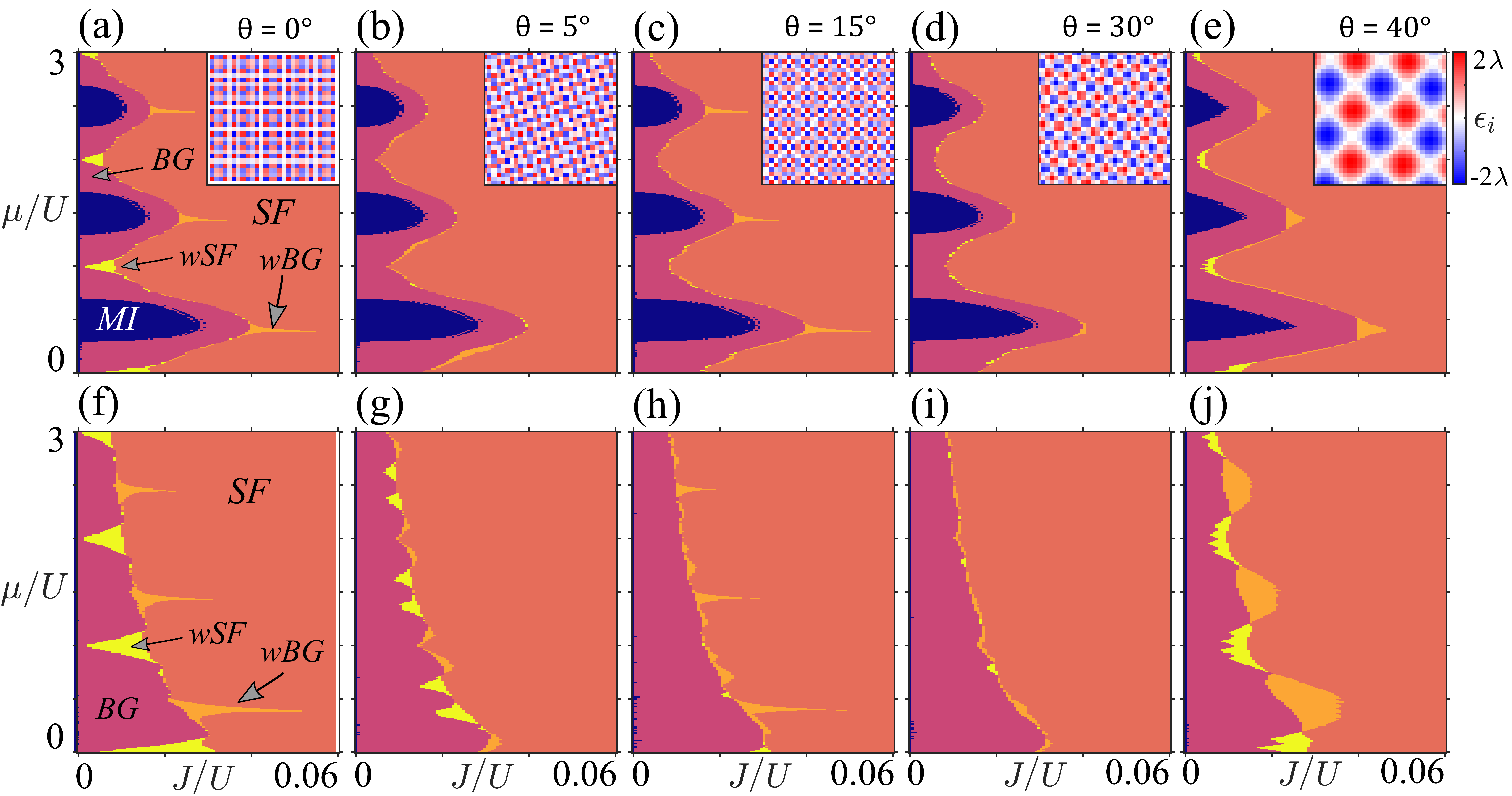}
	\caption{Phase diagrams of the tilted 2D AA model, for different rotation angles $\theta$. Here, we consider fixed modulation strengths of (a-e) $\lambda/U=0.15$ and (f-j) $\lambda/U=0.35$, with the inset figures representing small portions of the 2D AA potential. Each column represents a fixed rotation of (a,f) $\theta=0^{\circ}$, (b,g) $\theta=5^{\circ}$, (c,h) $\theta=15^{\circ}$, (d,i) $\theta=30^{\circ}$, (e,j) $\theta=40^{\circ}$. The BG will generally lose the lobe-like structure at finite $\theta$, but several unique features can still be observed due to patterns of weak modulation.}
	\label{figure_phMap_theta}
\end{figure*}

In Fig.~\ref{figure_phMap_theta}, we plot phase diagrams for two modulation strengths $\lambda/U$ over a range of rotation angles, with inset figures showing the tilted potential. Starting with small rotation angles in Figs.~\ref{figure_phMap_theta}, we observe that the tilted potential no longer stabilises weak modulation lines, but rather weakly modulated zig-zag patterns on the lattice. The overall structure of the phase diagrams is now far more reminiscent of those observed in randomly disordered systems, but now with the underlying phases possessing long-range order. We do observe regions where the ground state is a wSF or wBG for larger $\lambda/U$, but this is in relatively narrow regions for small $\theta$. This is evidence that weak modulation regions and mixed phases are playing a smaller role, but can still dominate the ground state properties in specific regions.

As $\theta$ increases, the AA potential will generally have less correlated regions of weak modulation, exaggerating the loss in observable effects and mixed phases. A special case is seen for $\theta=15^{\circ}$, where the extruding feature of the BG reappears on the phase diagram. By inspecting the AA potential in Fig.~\ref{figure_phMap_theta}(c), it can be seen that there are now diagonal lines of weak modulation, which will act as a barrier to superfluid percolation. SF clusters can also form on these diagonal lines, as shown in Figs.~\ref{figure_ord_map_R}(b,e). The SF or wSF phase is not stabilised around integer $\mu/U$, however, since the diagonal lines are not connected by bonds, and hence no percolation can occur through them. This again leads to the destruction of the lobe-like BG pattern on the phase diagram. By further increasing $\theta$ to $30^{\circ}$ in Figs.~\ref{figure_phMap_theta}(d,i), we observe similar properties to before, with the mixed phases no longer dominating large regions of the phase diagram.

For larger angles, e.g. $\theta = 40^{\circ}$, the mixed phases are actually enhanced as shown in Fig.~\ref{figure_phMap_theta}(e,j). This is due to the large rotation angles tending the potential towards a more uniform limit, with large regions of positive/negative on-site potential connected by extended flat regions; see the insert of Fig.~\ref{figure_phMap_theta}(e). We also note that for all considered phase diagrams here, no DW phases are found to be stable at finite tunnelling strengths.

From these results, we can see that properties of the mixed phases are strongly dependent on large, correlated domains of weak modulation. For $\theta = 0^{\circ}$, weakly modulated lines have a significant impact on the percolation of local SF clusters. When considering tilted potentials, we generally can no longer form regions of weak modulation across the entirety of the lattice. Even if smaller domains of weak modulation exist, these are connected by paths of stronger modulation, as is the case for the zig-zag patterns in Fig.~\ref{figure_phMap_theta}(b). For smaller rotations, we can still observe differences in the formation of mixed phases, which is a consequence of these weakly modulated patterns influencing local SF percolation.

\begin{figure}[t]
	\centering
	\includegraphics[width=0.6\linewidth]{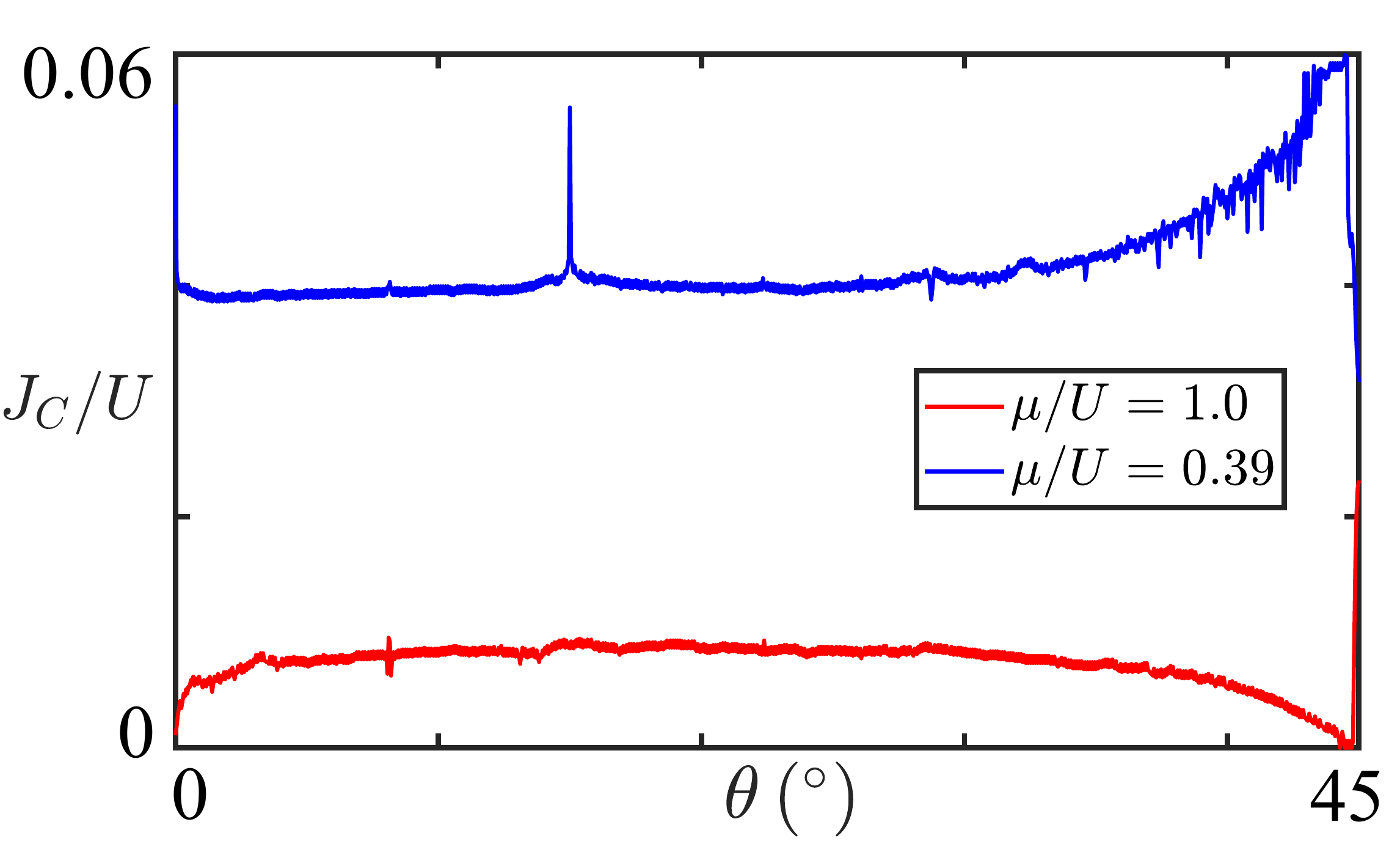}
	\caption{Plots of the critical behaviour at fixed chemical potentials for $\lambda/U=0.15$, showing the $J_C/U$ BG-SF critical point. The critical points remain stable for a large range of $\theta$. As $\theta \rightarrow 45^{\circ}$, the potential becomes more crystalline/uniform, which shifts the critical behaviour.}
	\label{figure_critJ_R}
\end{figure}

\subsubsection{Mixed Phases and Critical Points}

As we have seen, the study of $J/U$ critical points at $\mu/U=1$ and $\mu/U=0.39$ have been very important measures in the characterisation of mixed phases. In Fig.~\ref{figure_critJ_R}, we again plot the BG-SF critical points in $J/U$ for a range of tilt angles $\theta$. Here, we immediately observe several profound differences to what has been seen previously. First, the critical point at $\mu/U=1$ quickly rises for small rotations and oscillates around $J/U=0.008$ for the majority of $\theta$, which implies that the lobe-like structure of the BG will be partially destroyed. This is to be expected from the previous phase diagrams. Compared to the critical points with no rotations, there are no significant fluctuations in $J_C/U$ for the majority of tilt angles. As we approach $\theta=45^{\circ}$, the critical point will slowly decrease towards $0$, indicating the return of weakly modulated domains. At $\theta=45^{\circ}$, it can be seen that there is a sudden jump in the critical points. The reason for this can be inferred from the quasicrystalline distribution. Taking $\theta=45^{\circ}$, the 2D AA potential reduces to a flat distribution. The critical points at $\theta=45^{\circ}$ are therefore those of the disorder free system at an effective chemical potential of $\mu-2\lambda$.

At $\mu/U=0.39$, the BG-SF critical points also show significant differences. As before, the extruding feature effectively vanishes, and the critical point will remain stable at $J/U=0.04$ for most of the tilt angles. Curiously, however, at $\theta=15^{\circ}$, the extruding feature does return for a small range of tilt angles. The reason for this change is due to the existence of diagonal weakly modulated lines, as shown in Fig.~\ref{figure_phMap_theta}(c). As we approach the limit of $\theta=45^{\circ}$, the critical point slowly increases, before another sharp change when the 2D AA potential becomes more uniform. The tilted potentials substantially impact the formation and structure of mixed phases in the 2D AA model, but importantly do not completely destroy them.

\subsection{Skewed Potential with $\beta_1 \neq \beta_2$} \label{sc_5_3}

\subsubsection{Ground State Phase Diagrams}

\begin{figure*}[t]
	\centering
	\includegraphics[width=0.99\linewidth]{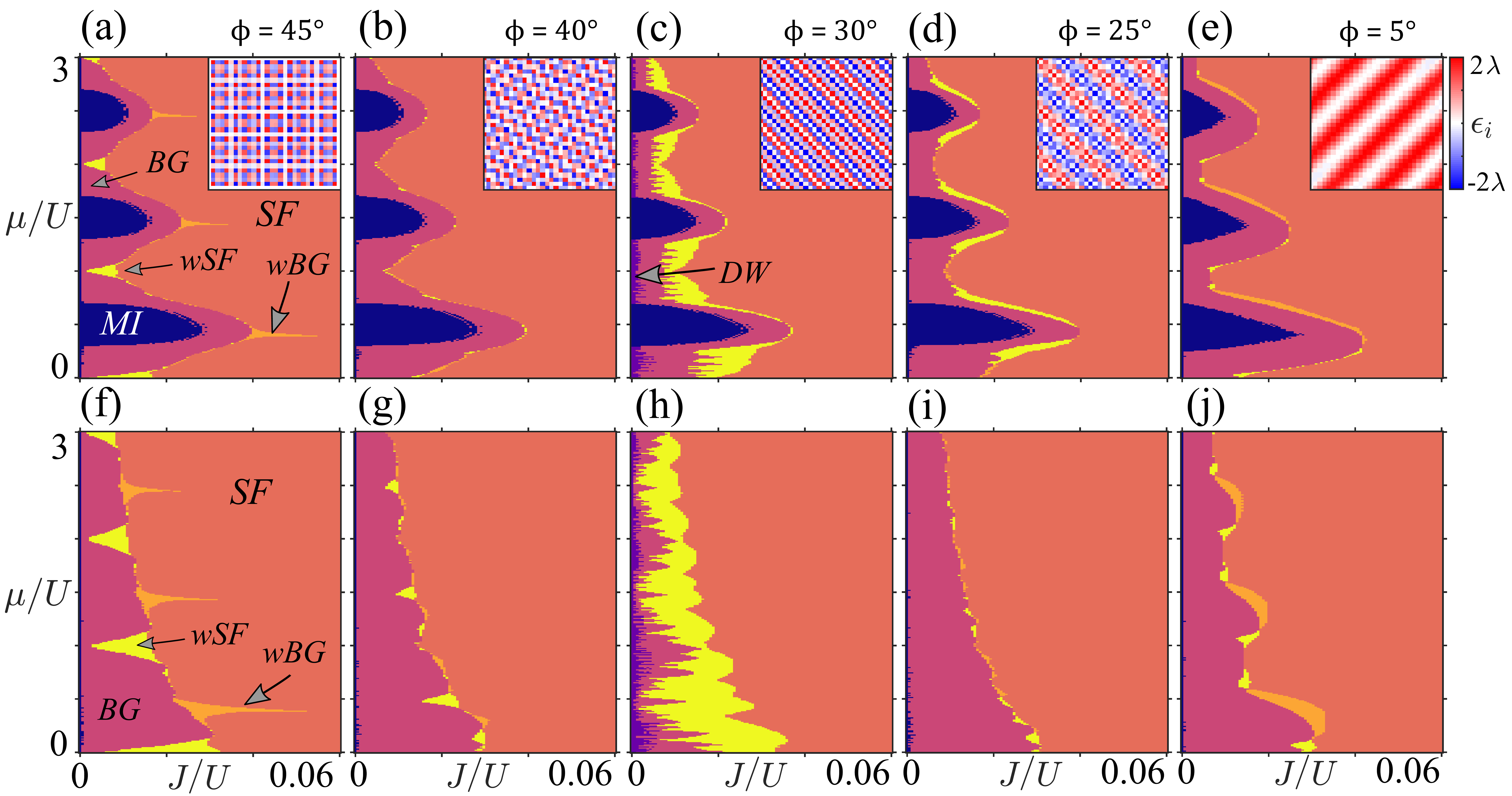}
	\caption{Phase diagrams of the skewed 2D AA model, for different skew angles $\phi$. Here, we consider fixed modulation strengths of (a-e) $\lambda/U=0.15$ and (f-j) $\lambda/U=0.35$, with the inset figures representing small portions of the 2D AA potential. Each column represents a skew angle of (a,f) $\phi=45^{\circ}$, (b,g) $\phi=40^{\circ}$, (c,h) $\phi=30^{\circ}$, (d,i) $\phi=25^{\circ}$, (e,j) $\phi=5^{\circ}$. The results share many similar properties those in Fig.~\ref{figure_phMap_theta}, with weakly modulated regions still giving rise to unique features on the phase diagrams. }
	\label{figure_phMap_phi}
\end{figure*}

In Fig.~\ref{figure_phMap_phi}, we plot phase diagrams and potentials for two modulation strengths $\lambda/U$, over a range of skew angles $\phi$. From these, we can see comparable results to what has been found with the tilted potentials. Starting with a skew angle of $40^{\circ}$ in Figs.~\ref{figure_phMap_phi}(b,g), the AA potential forms similar looking zig-zag patterns of weak modulation. This will generally destabilise the lobe-like structure of the BG, and remove the extruding features. As before, the wSF and wBG phases will become less pronounced, highlighting changes to the percolation of local SF/MI clusters. However, we do observe that the wSF and wBG phases are more robust for stronger AA potential strengths. This is also reflected in the structure of mixed phases, as seen in Figs.~\ref{figure_ord_map_L}(a,d).

For $\phi=30^{\circ}$ in Figs.~\ref{figure_phMap_phi}(c,h), we have a special case that has one rational and one irrational wavenumber, leading to a potential with lines of constant $\epsilon_i$ along the diagonals. Different clusters of phases will, therefore, only form across these diagonals, as per Figs.~\ref{figure_ord_map_L}(b,e). This results in a phase diagram that possesses quite a different structure from what has been seen previously. For instance, the BG now has multiple smaller lobes at certain $\mu/U$. The wSF shows even more profound differences due to the unique characteristics of this potential. Effectively, in this scenario, the wSF is a measure of a 1D MI-SF transition in the 2D structure. Furthermore, we also note that the number of unique values in the potential is comparable to the lattice side length due to the diagonal lines of constant $\epsilon_i$, leading to the appearance of DW states at finite tunnelling strengths around $J/U \approx 10^{-2}$.

At several of the intermediate skew angles, we also see phase diagrams comparable to Figs.~\ref{figure_phMap_theta}(d,i) and randomly disordered systems. Interestingly, the wSF can be stabilised across a larger set of chemical potentials, away from the BG domains. Finally, as $\phi \rightarrow 0^{\circ}$, the potential will again become more crystalline and uniform, similar to what was seen in Figs.~\ref{figure_phMap_theta}(e,j) but now with a diagonal, effectively 1D, form.

\subsubsection{Mixed Phases and Critical Points}

\begin{figure}[t]
	\centering
	\includegraphics[width=0.6\linewidth]{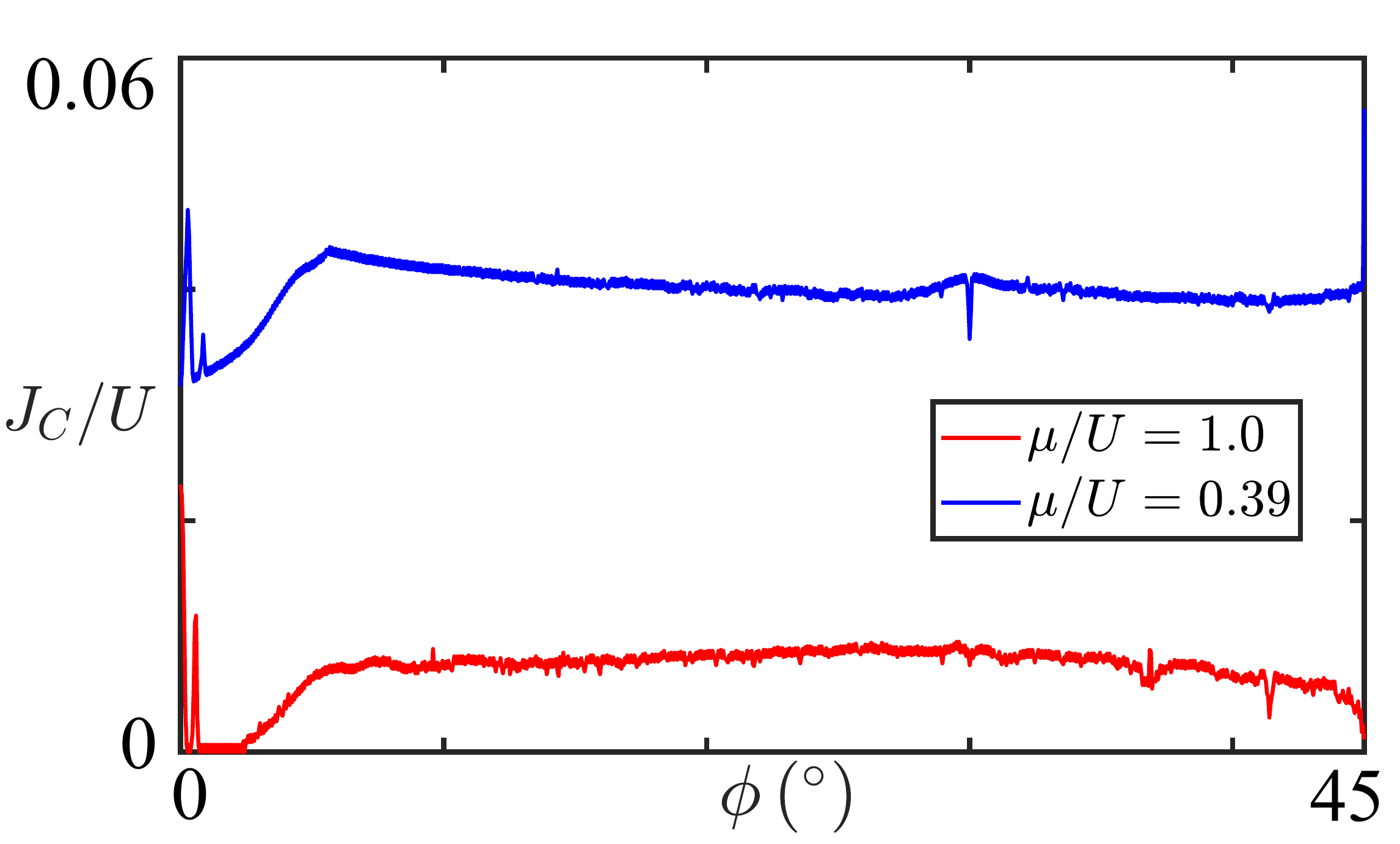}
	\caption{Plots of the critical behaviour at fixed chemical potentials for $\lambda/U=0.15$, showing the $J_C/U$ BG-SF critical point. The critical points are again stable for a large range of $\phi$. As $\phi \rightarrow 0^{\circ}$, the potential becomes more crystalline/uniform, which shifts the critical behaviour.}
	\label{figure_critJ_L}
\end{figure}

Here, we will again plot how certain $J/U$ critical points are influenced across a full range of $\phi$ in Fig.~\ref{figure_critJ_L}. From these results, we observe some very immediate similarities to what was seen with the tilted potentials. Moving away from $\phi=45^{\circ}$, the lobe-like structure and extruding features of the BG are again destroyed, and the critical points remain stable across the majority of $\phi$. This is to be expected, as the skewed potentials will also no longer stabilise precise lines of weak modulation throughout the lattice. As we approach $\phi=0^{\circ}$, both critical points will start decreasing towards a fixed value, before a sudden jump. This behaviour is again due to the emergence of crystalline properties within this interval, with a flat distribution of on-site energies at $\phi=0^{\circ}$. We finally note that both the skewed and tilted potentials share many similar properties in their critical behaviour and features on phase diagrams. This is due to the fact that both kinds of potentials can stabilise similar kinds of weakly modulated zig-zag patterns on the lattice, which allows for the formation of mixed phases.

\section{Conclusions} \label{sc_6}

In summary, we have shown the presence of intriguing mixed phases in the many-body 2D AA model. The AA potential is markedly distinct from randomly disordered systems, as it may permit the formation of large, correlated domains that possess weak modulation. These domains can then act as barriers to macroscopic superfluidity and insulation, with the long-range order playing a key role in the percolation of MI and SF clusters throughout the lattice. This can dramatically shift critical behaviour, with the most striking feature being the appearance of BG lobes with sharp, wBG extrusions, which we have now confirmed to exist across a wide range of quasicrystalline distributions. At integer $\mu/U$, the insulating behaviour of the BG is destroyed through the percolation of small SF clusters, resulting in a wSF phase. We have also linked the appearance of these features to changes in the underlying structure of mixed phases. In particular, we find that local SF clusters are either the majority or minority of the phase at irrational wavenumbers, due to localisation across precise lines of weak modulation.

By considering more general AA potentials, we have also shown that unique properties of the mixed phases can still be observed, provided that there are smaller domains of weak modulation. If these domains do not exist in the 2D AA potential, we actually find results comparable to those in randomly disordered systems, but with local structures now possessing long-range order.

Furthermore, we have also studied the importance of long-range variations within the on-site potential in stabilising precise regions of mixed phases. At the superlattice limits of the 2D AA model, there is a small number of unique values in the energy distribution, which implies the presence of both short- and long-range order. As a result, the percolation and transition of mixed phases will become more correlated and uniform, leading to greater shifts in critical behaviour. On a global scale, phase boundaries can dramatically change, and DW phases at finite tunnelling strengths can also form. On the other hand, when we consider rational wavenumbers that approach quasicrystalline distributions, the influence of long-range variations in on-site energy becomes clear. The number of unique values in the energy distribution will be comparable to the total number of lattice sites, leading to the formation of much fewer regions of weak modulation that stabilise SF percolation. The number of unique DW states will also be vast, and possess a strong sensitivity to particle number fluctuations.


\ack
D.J., acknowledges support from EPSRC CM-CDT Grant No. EP/L015110/1. Work at the University of Strathclyde was supported by the EPSRC Programme Grant DesOEQ (EP/P009565/1), and the EPSRC Quantum Technologies Hub for Quantum Computing and Simulation (EP/T001062/1).

\appendix
\newpage

\section{Mixed Phases Towards the Crystalline Limit} \label{sc_ap1}

\begin{figure}[t]
	\centering
	\includegraphics[width=0.85\linewidth]{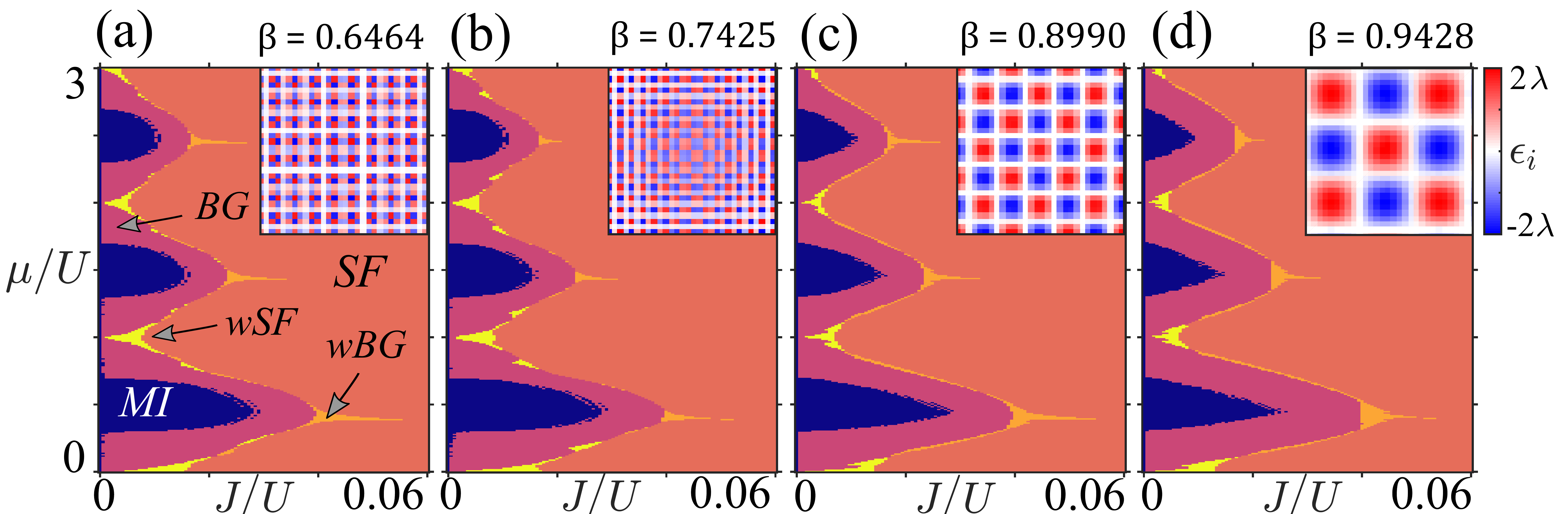}
	\caption{Phase diagrams of the 2D AA model for different irrational wavenumbers $\beta$, at a fixed modulation strength of $\lambda/U=0.15$. The inset figures represent small portions of the 2D AA potential. We also express the wavenumber to 4 decimal places, for comparison with Fig.~\ref{figure_uniqueV}. The exact wavenumbers are given as (a) $\beta=\frac{2\sqrt{2}-1}{2\sqrt{2}}$, (b) $\beta=\frac{21}{20\sqrt{2}}$, (c) $\beta=\frac{7\sqrt{2}-1}{7\sqrt{2}}$, (d) $\beta=\frac{2\sqrt{2}}{3}$. The BG retains a lobe-like structure across a large range of irrational wavenumbers due to the presence of weakly modulated lines. When $\beta \rightarrow 1$, the potential takes a more regular and crystalline form.}
	\label{figure_phMap_beta_2}
\end{figure}

In this appendix, we will briefly consider the influence of mixed phase as we approach a crystalline limit of the 2D AA model. We will take $\beta_1=\beta_2 \equiv \beta$, and plot full phase diagrams towards $\beta=1$. In Fig.~\ref{figure_phMap_beta_2}, we plot these regions for different cases of $\beta$, with the insets figures showing a small portion of the 2D AA potential for reference. Here, we generally see very similar behaviour and properties to what is observed in Fig.~\ref{figure_phMap_md}, with the BG retaining its lobe-like structure for a wide range of wavenumbers. The wSF remains localised around integer chemical potentials, highlighting the stability of weak modulation lines in forming local SF clusters. Furthermore, the wBG remains localised to the sharp extruding feature of the BG. As we approach $\beta = 1$ in Figs.~\ref{figure_phMap_beta_2}(c,d), the potential will begin to take a larger and more regular form on the lattice. Despite this, the BG will still maintain its lobe-like appearance, and may even further destabilise the MI for $\beta \rightarrow 1$, giving rise to a triangular structure on the phase diagram. We also observe that the extruding features emerging from the overall BG lobe are persistent for a range of $\beta$, with the wBG forming across similar regions throughout.



\providecommand{\newblock}{}

\end{document}